\documentclass[manuscript]{aastex}

\shorttitle{M6.6-class Solar Flare in NOAA AR 11158}
\shortauthors{Toriumi et al.}

\begin{document}


\title{Magnetic Systems
  Triggering the M6.6-class Solar Flare
  in NOAA Active Region 11158}


\author{Shin Toriumi\altaffilmark{1} and Yusuke Iida\altaffilmark{2}}
\affil{Department of Earth and Planetary Science, University of Tokyo, Hongo, Bunkyo-ku, Tokyo 113-0033, Japan}
\email{toriumi@eps.s.u-tokyo.ac.jp}

\author{Yumi Bamba, Kanya Kusano\altaffilmark{3}, and Shinsuke Imada}
\affil{Solar-Terrestrial Environment Laboratory, Nagoya University, Furo-cho, Chikusa-ku, Nagoya, Aichi 464-8601, Japan}

\and

\author{Satoshi Inoue}
\affil{School of Space Research, Kyung Hee University, Yongin, Gyeonggi-do, 446-701, Republic of Korea}

\altaffiltext{1}{JSPS Research Fellow}
\altaffiltext{2}{present address: Institute of Space and Astronautical Science, Japan Aerospace Exploration Agency, Chuo-ku, Sagamihara, Kanagawa 252-5210, Japan}
\altaffiltext{3}{Japan Agency for Marine-Earth Science and Technology (JAMSTEC), Kanazawa-ku, Yokohama, Kanagawa 236-0001, Japan}




\begin{abstract}
We report a detailed event analysis
on the M6.6-class flare
in the active region (AR) NOAA 11158
on 2011 February 13.
AR 11158,
which consisted of two major emerging bipoles,
showed prominent activities
including one X- and several M-class flares.
In order to investigate
the magnetic structures
related to the M6.6 event,
particularly the formation process
of a flare-triggering magnetic region,
we analyzed multiple spacecraft observations
and numerical results
of a flare simulation.
We observed that,
in the center of this quadrupolar AR,
a highly sheared polarity inversion line (PIL)
was formed
through proper motions
of the major magnetic elements,
which built a sheared coronal arcade
lying over the PIL.
The observations lend support
to the interpretation that
the target flare was triggered
by a localized magnetic region
that had an intrusive structure,
namely a positive polarity
penetrating into a negative counterpart.
The geometrical relationship
between the sheared coronal arcade
and the triggering region
was consistent
with the theoretical flare model
based on the previous numerical study.
We found that
the formation of the trigger region
was due to a continuous accumulation
of the small-scale magnetic patches.
A few hours before the flare occurrence,
the series of emerged/advected patches
reconnected with a preexisting fields.
Finally, the abrupt flare eruption
of the M6.6 event started
around 17:30 UT.
Our analysis suggests that,
in a triggering process
of a flare activity,
all magnetic systems
of multiple scales,
not only the entire AR evolution
but also the fine magnetic elements,
are altogether involved.
\end{abstract}






\section{Introduction\label{sec:intro}}

It is now widely believed that
solar flares
and resultant various eruptions
are massive explosions
with the help of magnetic reconnection.
On this issue,
many observational and theoretical studies
have been carried out.
Large-scale sunspot motions produce
strong magnetic shear \citep{hag84}
and sharp magnetic gradient \citep{schr07}.
The nonpotentiality
in an active region (AR)
is thus achieved
and, through this process,
free energy is stored
in the corona.
For starting the release of the stored energy,
the existence of driving mechanism
is thought to be necessary.
Such mechanism includes
emerging flux \citep{hey77,fey95,che00},
converging motion
and flux cancellation
\citep{van89,moo01}
at the polarity inversion line (PIL).
Therefore,
we can understand flares
as the phenomena
that the stored
magnetic free energy
in the corona
is released through instability
and magnetic reconnection,
which are triggered
by some mechanism.
However, the causal interaction
among the magnetic structures
and the resultant process of flares
still remained unclear.

\begin{figure}
  \includegraphics[scale=0.9,clip]{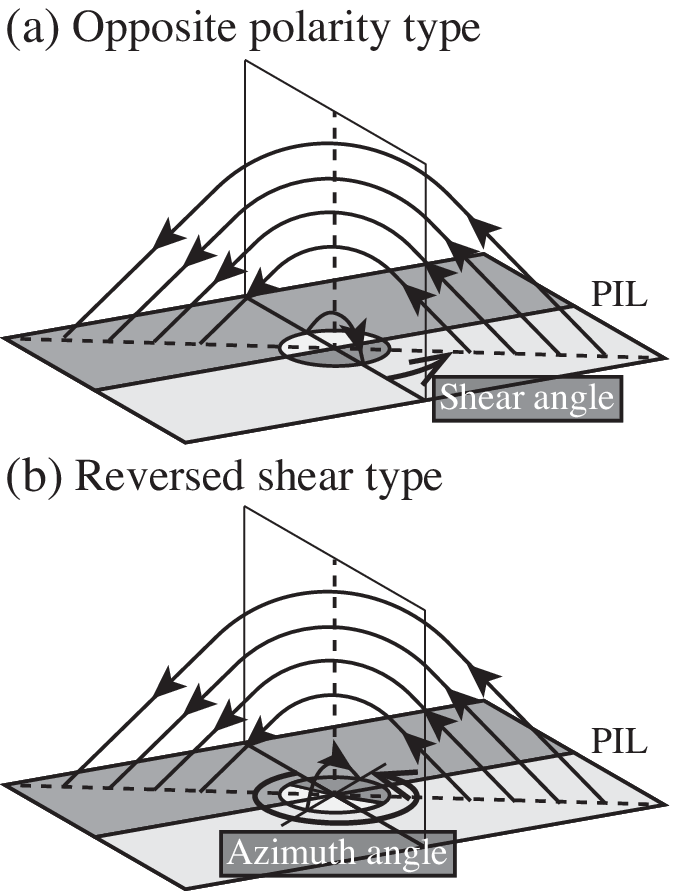}
  \caption{Schematic illustration
    of the two different types
    of flare onset
    suggested by \citet{kus12}:
    (a) opposite polarity (OP) type
    and (b) reversed shear (RS) type.
    In both cases,
    coronal arcade fields
    lying over the PIL
    and the local triggering field
    (emerging flux)
    are shown by curved thin arrows,
    while photospheric magnetic fields
    are indicated
    by lighter (positive)
    and darker (negative) hatches.
    The shear angle of the overlying arcade
    and the azimuth angle of the triggering field
    are the parameters
    that characterize the magnetic configuration.
    Both angles are measured counterclockwise
    from the axis normal to the PIL
    (thick arrows).
   }
  \label{fig:kus12}
\end{figure}

Recently,
a systematic study of
three-dimensional magnetohydrodynamic (MHD) simulations
in terms of
the flare-triggering mechanisms
has been performed
by \citet{kus12}.
They modeled the preexisting coronal arcade
overlying the PIL of an AR
and a triggering field
injected at the bottom boundary.
By varying the shear angle of the coronal arcade
and the azimuth angle of the
injected flux,
they systematically surveyed
the conditions of the flare onset.
As a result, it is found that
there are two different types
for the flare onset
depending on
the azimuth angle
of the injected flux
(Figure \ref{fig:kus12}).
One is the opposite-polarity (OP) case,
where the injected bipole on the PIL
has an opposite orientation
to the overlying arcade
(Panel a),
and the other is
the reversed-shear (RS) case,
where the azimuth of the bipole
is sheared in a reversed sense
to the arcade
(Panel b).
In addition,
\citet{kus12} analyzed
the observational data
of two major flares,
an X3.4-class flare
on 2006 December 13
in NOAA AR 10930
and an M6.6 flare
on 2011 March 13
in AR 11158,
and compared these observations
with their numerical results.
They found that
the magnetic configuration
and the location
of preflare brightening
in the X3.4 flare
are suitable
for the OP-type case,
while those in the M6.6 flare
are for the RS case.

As for observational studies,
the next targets should be
the formation process
of flare trigger
and its interaction
with coronal arcades
that drives a resultant eruption.
In the calculations by \citet{kus12},
the flare trigger
was simply given
in the form of
a local flux emergence
at the bottom boundary
that corresponds to
OP- or RS-type configuration.
In the actual Sun,
however,
such a magnetic configuration
could be achieved
through
a variety of dynamic processes
including
small flux emergence.
Recent progress of
spatially- and temporally-resolved observations
will contribute
to the further understanding
of trigger development
in the preflare phase
as well as
the whole AR evolution
and the consequent eruption.

In this paper,
we present a detailed event analysis
on the M6.6-class flare
occurred on 2011 February 13
in NOAA AR 11158,
which was briefly reported
in \citet{kus12}.
Here,
we use
observational data of this AR
taken by multiple spacecrafts,
along with
the nonlinear force-free field (NLFFF) extrapolation
and numerical results
of a flare simulation.
The targets of this paper are
(1) the overall development
of the magnetic fields
in AR 11158
before the M6.6 flare,
(2) the formation process
of the flare-triggering region
that initiates the M flare,
and (3) the evolution
of the M flare
in the main phase.
The rest of the paper
proceeds as follows.
In Section \ref{sec:observation},
we introduce observations
of AR 11158
and data reduction processes.
The obtained
magnetic field configurations
and the formation of the flare trigger
are shown in Sections
\ref{sec:arcade} and \ref{sec:trigger},
respectively.
Then,
in Section \ref{sec:flare},
we compare observational results
of the M flare
with numerical simulations
by \citet{kus12}.
Discussion and summary
are presented
in Sections \ref{sec:discussion}
and \ref{sec:summary},
respectively.

\section{Observations and Data Reductions
  \label{sec:observation}}

\subsection{AR 11158 and the M6.6-class Flare}

\begin{figure}
  \includegraphics[scale=1.,clip]{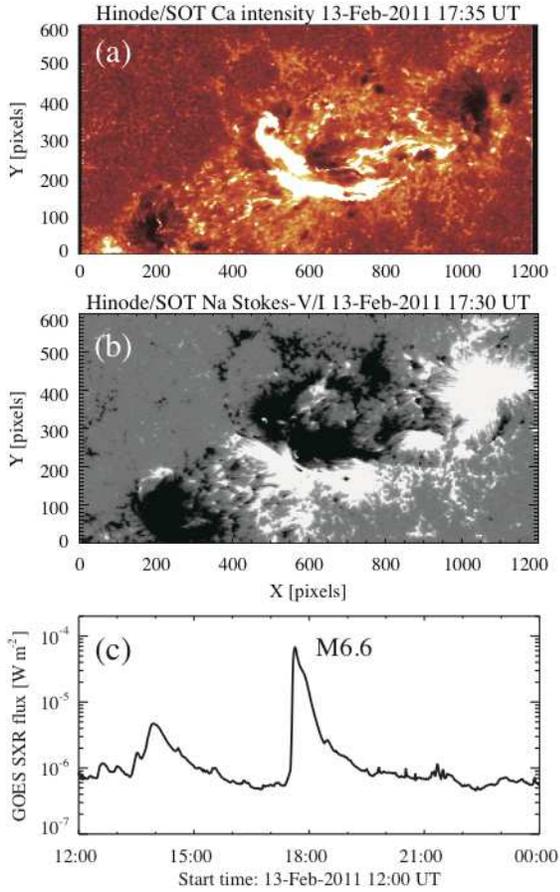}
  \caption{
    The M6.6-class flare
    in NOAA AR 11158.
    (a) Ca intensity map
    at 17:35 UT
    and (b) Na Stokes-V/I image
    at 17:30 UT
    on 2011 February 13,
    observed by {\it Hinode}/SOT.
    (c) {\it GOES}-15 soft X-ray flux
    of $1.0$--$8.0\ {\rm \AA}$ channel
    (1-min cadence).
   }
  \label{fig:flare}
\end{figure}

The target flare
of this study
is the M6.6-class event
on 2011 February 13
in NOAA AR 11158
(see Figure \ref{fig:flare}).
This AR appeared on
the southern hemisphere
on 2011 February
and was composed of
two major emerging bipoles.
Since the whole evolution
of this AR,
from its birth to the flares,
occurred in the near side
of the Sun,
we can particularly investigate
the evolution history
of the magnetic fields
related to the flares
from its earliest stage.
Above the highly sheared PIL
located at the center
of this quadrupolar region,
many flares
including one X- and some M-class events
were observed.
The M6.6 flare broke out
around 17:30 UT
on February 13,
which can be seen
in the {\it Geostationary Operational Environmental Satellite}
({\it GOES}) soft X-ray flux
($1.0$--$8.0\ {\rm \AA}$ channel)
in Figure \ref{fig:flare}c.
The M6.6 event here was about 32 hr
before the X2.2 flare,
the first X-class flare
of Solar Cycle 24,
which occurred on the same PIL
\citep{schr11}.

In \citet{kus12},
they identified
a flare-triggering region
of the M6.6 event
on February 13
and found that
the triggering region
had an RS configuration.
The formation of this trigger,
however,
still remained unclear.
In this study,
we focus on the magnetic fields,
especially on
the formation
of the trigger region,
by using observational data
introduced below.

\subsection{{\it Hinode}/SOT Data}

The {\it Hinode} satellite \citep{kos07}
tracked NOAA AR 11158
from 09:57 UT, February 12,
to 09:01 UT, February 19.
During this period,
the Filtergram (FG)
of the Solar Optical Telescope
\citep[SOT:][]{tsu08}
on board {\it Hinode} obtained
circular polarization (CP) data
of \ion{Na}{1} D$_{1}$ line (5896 \AA),
shifted by $140\ {\rm m\AA}$
from the line center,
which is called Stokes-V/I images hereafter,
and intensity data
of \ion{Ca}{2} H line (3968.5 \AA).
The time cadence
is 5 min for each data,
and the original field of view (FoV) is
$225.3''\times 112.6''$ for Stokes-V/I
and $183.2''\times 108.5''$ for Ca data,
respectively.
The spatial sampling is
$0.16''$ for Stokes-V/I images
and $0.108''$ for Ca images.
Both data are calibrated
through {\tt fg\_prep} procedure
included in {\it SolarSoftWare} (SSW) package
for dark-current subtraction and flat fielding.
Then, by taking a cross-correlation
between two consecutive images,
we reduce
small spatial fluctuations.
Finally,
Ca images are enlarged
to fit the size
of the Stokes-V/I data
and structures
in these images
are compared.

The Spectro-Polarimeter (SP) of the SOT
takes spectrum profiles
of two magnetically sensitive \ion{Fe}{1} lines
at 6301.5 and 6302.5 \AA.
The SP scan data
obtained in the observational period
on February 13
have a time cadence
of 90--120 min
and a spatial sampling of $0.32''$.
The FoV is basically $164''\times 164''$;
some scans have a lack of spectral data.
In this study
we use the SOT/SP level2 data,
which are the outputs
from the inversions
using the Milne-Eddington gRid Linear Inversion Network
(MERLIN) code \citep{lit07}.

\subsection{{\it SDO}/AIA and HMI Data
  \label{sec:sdo}}

The Atmospheric Imaging Assembly
\citep[AIA:][]{lem12}
on board the {\it Solar Dynamics Observatory} ({\it SDO})
continuously observes
the coronal dynamics
of the whole solar disk
at multiple wavelengths.
In this study
we use the tracked cutout data
of AR 11158
with a time cadence of $12\ {\rm s}$
and a pixel size of $0.60''$,
which are taken from
the {\it SDO} AIA Get Data page.
\footnote{http://www.lmsal.com/get\_aia\_data/}
We also use the magnetogram
of the Helioseismic and Magnetic Imager
\citep[HMI:][]{sche12,scho12}
of the same FoV
with a time cadence of $45\ {\rm s}$
and a pixel size of $0.50''$,
along with the vector magnetic field data
with a cadence of 12 min,
which is cut out
from the full disk data
using HARP
\citep{tur02,tur10}.
\footnote{http://jsoc.stanford.edu/jsocwiki/VectorDataReference}

\subsection{Nonlinear Force-Free Field Data}

NLFFF extrapolation
is also applied to the vector field data
taken by {\it SDO}/HMI
(Section \ref{sec:sdo})
in order to investigate
the three-dimensional magnetic structure.
In this study,
the MHD relaxation method
developed by \citet{ino13}
is employed to reconstruct
the force-free field.
The initial and boundary conditions for the
iterative calculation are given
by a three-dimensional potential field
and the preprocessed vector magnetogram.
The detailed procedure will be
explained in Appendix \ref{app:nlfff}.

The NLFFF model covers the domain
$L_{x} \times L_{y} \times L_{z}
= 184.32 \times 184.32 \times 184.32\ {\rm Mm}^{3}$,
which is resolved
by $128 \times 128 \times 128$ grids.
The bottom boundary condition
is obtained from
a $4 \times 4$ binning
of the original vector magnetogram
of $512 \times 512$ pixels.
The grid size of the NLFFF
corresponds to about $1688\ {\rm km}$,
i.e., about $2.3''$.

\section{Magnetic Field Configuration: Preflare State
  \label{sec:arcade}}

In this section,
we first describe
the overall development
of the magnetic field
in the preflare stage.
Then we focus on
the magnetic configuration
of the possible candidate
for the flare trigger
of the M6.6-class flare,
which
was localized
in the center of this AR.

\subsection{Initial Phase and the Formation of Sheared PIL}

\begin{figure}
  \includegraphics[scale=1.,clip]{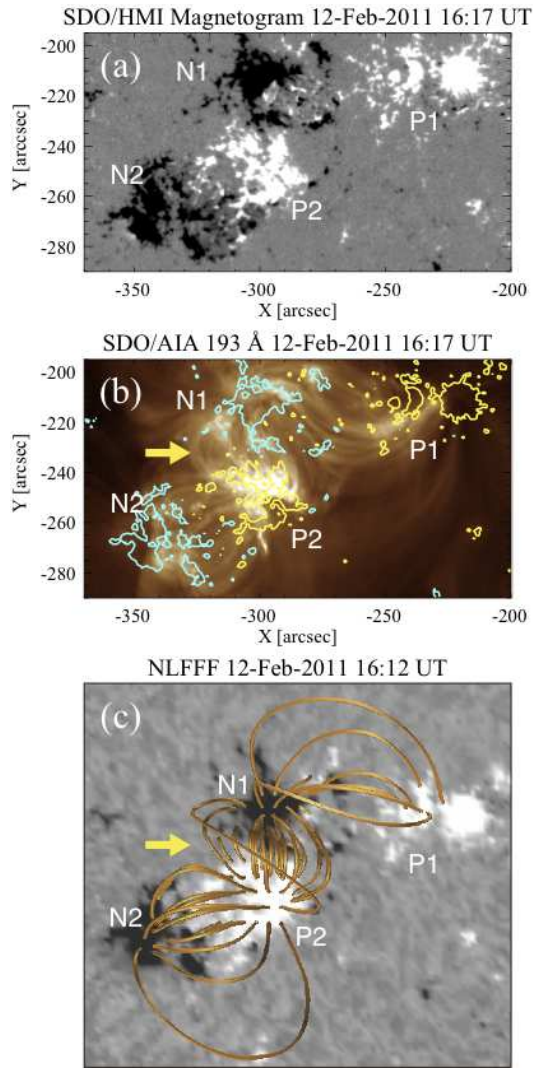}
  \caption{(a) {\it SDO}/HMI magnetogram
    of NOAA AR 11158
    taken at 16:17 UT
    on 2011 February 12,
    $\sim$1 day before
    the M6.6-class flare.
    The grayscale saturates
    at $\pm 200\ {\rm G}$
    and the axes are
    in arcseconds
    from disk center.
    Numbers representing
    two bipolar pairs
    ({\sf P1}--{\sf N1} and {\sf P2}--{\sf N2})
    are overplotted.
    (b) {\it SDO}/AIA 193 \AA\ image
    at the same time as (a).
    Contour levels of $\pm 200\ {\rm G}$
    are indicated
    as yellow and  turquoise lines.
    (c) NLFFF
    calculated from HMI magnetogram.
    Arrows in Panels (b) and (c)
    show the coronal arcade
    connecting {\sf N1}
    and {\sf P2}.
   }
  \label{fig:arcade}
\end{figure}

In its initial phase,
AR 11158 was composed of
two major emerging fluxes.
Figure \ref{fig:arcade}a
shows the {\it SDO}/HMI magnetogram
of this AR
at 16:17 UT
on February 12,
about 1 day before
the M6.6-class flare.
The first bipole
({\sf P1}--{\sf N1}) appeared
on the southern hemisphere
on February 9,
while the second pair ({\sf P2}--{\sf N2})
appeared on 10.
Both pairs,
of which the field strengths
were larger than
$1000\ {\rm G}$,
gradually
separated from each other
since their births,
with increasing flux
sourced from
a series of minor emergences.

Figure \ref{fig:arcade}b
shows the {\it SDO}/AIA 193 \AA\ image
at the same time as Panel (a).
This figure indicates that
there were coronal arcades
connecting magnetic patches
{\sf N1} and {\sf P2}
(indicated by an arrow)
as well as
arcades connecting {\sf P1}--{\sf N1}
and {\sf P2}--{\sf N2}.
Such arcades
are also seen
in the NLFFF map,
in Panel (c),
which is calculated from
the HMI magnetogram
as a bottom boundary condition.
Here one may speculate that
the original loops
of {\sf P1}--{\sf N1}
and {\sf P2}--{\sf N2}
conflicted over the PIL
between {\sf N1} and {\sf P2}
and made new loops of {\sf N1}--{\sf P2}
via magnetic reconnection.
The other half of
the reconnected loops
bridging between {\sf P1}--{\sf N2}
may lie
above the entire AR,
which is not seen in the NLFFF map
in Panel (c).

During the evolution of this AR,
the southern positive patch {\sf P2}
continuously moved to the west,
while the northern negative patch {\sf N1}
drifted to the east.
Therefore,
the coronal arcade {\sf N1}--{\sf P2}
was continuously sheared
in the counterclockwise direction
throughout the whole evolution process
(right-handed shear),
which can be seen
by comparing Figures \ref{fig:arcade}(b) and (c)
and Figures \ref{fig:preflare}(g) and (h).
Such a continuous shearing
strongly indicates
the storage of free energy
in the corona
above the PIL
between {\sf N1} and {\sf P2}.
On the contrary,
{\sf P1} and {\sf N2} moved slowly
and remained
almost at the same positions
from February 14.
Therefore,
the large-scale evolution in this AR
can be described as
the inner two polarities {\sf N1} and {\sf P2}
moved rapidly
as if they merged
into outer two polarities {\sf N2} and {\sf P1},
respectively,
creating a highly sheared PIL
between {\sf N1} and {\sf P2}.

\subsection{Intrusive Magnetic Structure on the PIL}

\begin{figure}
  \includegraphics[scale=1.,clip]{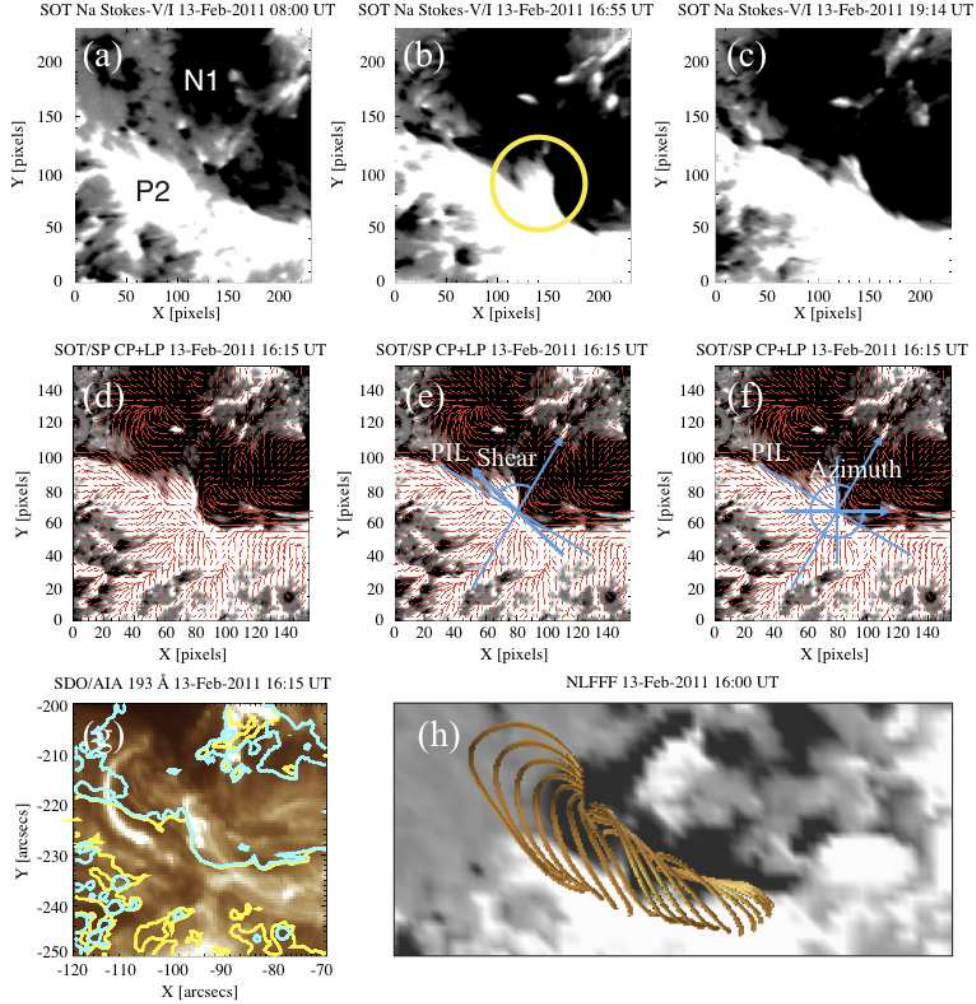}
  \caption{
    (a--c)
    Time-development
    of the PIL
    between {\sf N1} and {\sf P2}
    in Na Stokes-V/I maps
    taken by {\it Hinode}/SOT.
    White and black represent
    positive and negative polarities,
    respectively.
    One can see that
    the intrusive magnetic structure
    is formed along the PIL
    before the M6.6-class flare
    starting from 17:30 UT
    (yellow circle).
    (d--f) {\it Hinode}/SOT SP
    circular polarization (CP) map
    before the flare.
    Linear polarization (LP)
    that represents
    transverse field
    is overplotted
    as red small bars.
    The time shown
    is when the SP scan slit
    comes to the center
    of the field of view.
    Auxiliary lines for measuring
    the shear and azimuth angles
    are overplotted
    in Panels (e) and (f).
    (g) {\it SDO}/AIA 193 \AA\ image
    at the same time as (d--f),
    overplotted with
    field strength contours
    of $\pm 200\ {\rm G}$.
    Coronal arcades are
    across the PIL
    with a shear,
    which is also seen
    in (h) the NLFFF map.
   }
  \label{fig:preflare}
\end{figure}

As the AR developed,
{\sf N1} and {\sf P2} came
close to each other
and the magnetic gradient
across the PIL {\sf N1}--{\sf P2} increased.
Figures \ref{fig:preflare}(a--c)
are SOT Na Stokes-V/I images,
showing the temporal evolution
of this PIL.
In Panel (a),
at 08:00 UT on February 13,
there still remained
a polarity gap
between {\sf N1} and {\sf P2} patches.
In Panel (b),
about 30 min before
the M6.6-class flare took place,
however,
the gap had been filled
with a positive polarity.
Here one may find that
an intrusive magnetic structure,
where the positive polarity {\sf P2}
penetrates into
the negative polarity {\sf N1},
has been formed on the PIL
(yellow circle).
This magnetic structure
weakened
after the flare finished
in Panel (c).
\citet{kus12} reported that,
from the comparison
with flare simulations,
this intrusive structure
is the flare-triggering region
of the M6.6 flare.

Figures \ref{fig:preflare}(d--f)
are the SOT/SP map
taken at 16:15 UT
on February 13.
These images show
vertical and transverse fields
at the photosphere,
indicating the precise magnetic configuration
around the intrusive structure.
In Panel (d),
the transverse field (red bars)
was highly sheared
and was almost along the PIL
because of the
counterclockwise motion
of the both polarities
(right-handed shear).
The 193 \AA\ image
in Panel (g)
and the NLFFF map
in Panel (h)
also show that
the overlying coronal arcade
connecting {\sf N1} and {\sf P2}
was highly sheared
at this time.
This sheared PIL
had a length of
more than $20\ {\rm Mm}$.
To estimate the shear angle
of the overlying arcade
and the azimuth
of the local magnetic structure
from the PIL,
we overplotted 
auxiliary lines
in Panels (e) and (f).
Following Figure \ref{fig:kus12},
we measured
the shear and azimuth angles,
counterclockwise
from the normal of the PIL.
Here the shear is evaluated
from the transverse field,
while the azimuth is
from the western edge
of the intrusive structure.
The measured shear angle
is $\sim 80^{\circ}$
and the azimuth
ranges for $270^{\circ}$--$300^{\circ}$.

From the observations
in this section,
we see that the large-scale coronal arcade
above the PIL
satisfies the sufficient conditions
for large flares
proposed by \citet{hag90}.
That is,
(1) the shear angle of the coronal arcade
is as much as $\sim 80^{\circ}$,
(2) the field strength is large enough
($\ge 1000\ {\rm G}$),
and (3) the length of the PIL
is large enough ($\ge 10\ {\rm Mm}$).

\section{Formation of the Intrusive Structure: Flare Trigger
  \label{sec:trigger}}

\begin{figure}
  \includegraphics[scale=0.8,clip]{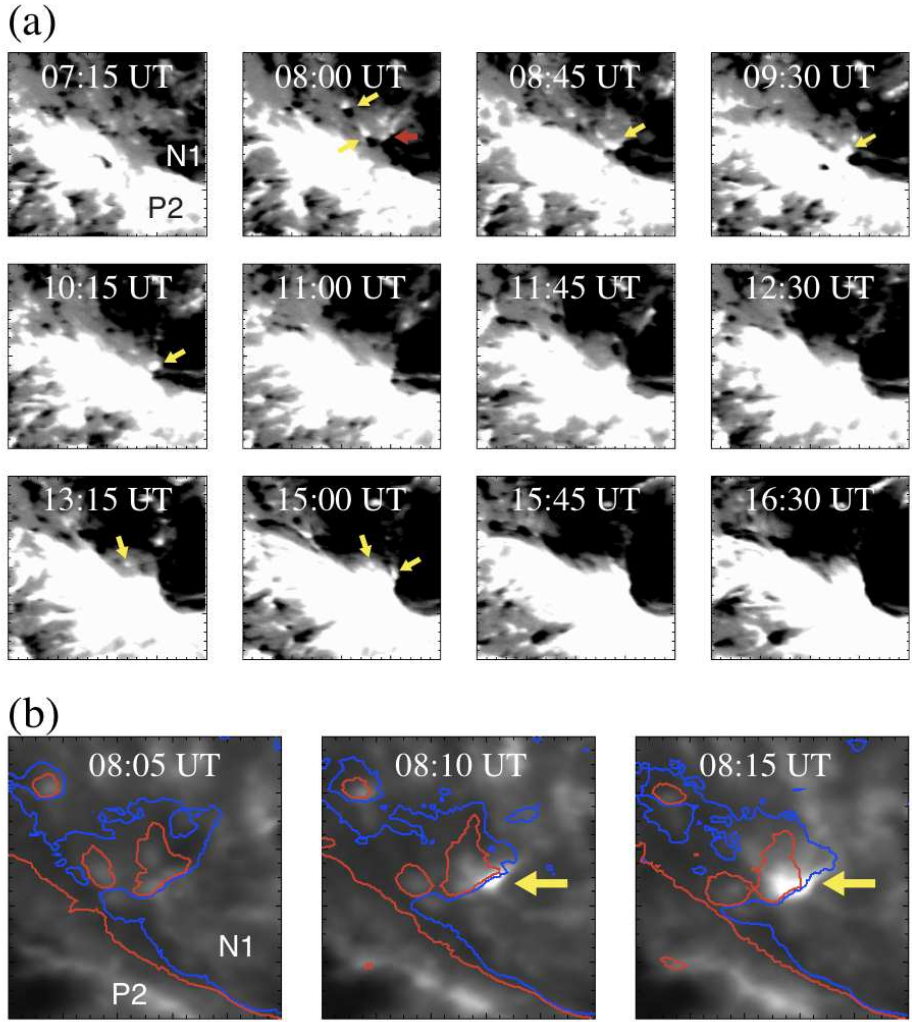}
  \caption{(a) SOT Na Stokes-V/I maps
    from 07:15 to 16:30 UT,
    2011 February 13,
    indicating the formation
    of the intrusive structure
    on the PIL between {\sf N1} and {\sf P2}.
    Arrows show the small-scale magnetic bipoles
    that collide into the preexisting polarities.
    (b) SOT Ca images
    of the collision event
    indicated by a red arrow
    in Panel (a).
    Red and blue contours
    show the data number (DN)
    of the circular polarization
    $=\pm 20$, respectively.
    Arrows indicate
    the Ca brightenings
    at the collision site
    of the two polarities.
   }
  \label{fig:trigger}
\end{figure}

Figure \ref{fig:trigger}a is
the SOT Na Stokes-V/I maps
around
the intrusive magnetic structure
(the flare-triggering region)
from 07:15 to 16:30 UT
on February 13.
Note that
there was a temporal gap
in the SOT data
between 13:15 and 15:00 UT,
since,
due to the telemetry limitation
of the {\it Hinode} satellite,
the SOT data in this time range
was overwritten.
As can be seen from Figure \ref{fig:trigger}a,
initially at 07:15 UT,
there still was a wide gap
on the PIL
between the two major polarities
({\sf N1} and {\sf P2})
and here the intrusive structure
was not evident.
As time went on,
a series of small-scale magnetic bipoles
emerged at the PIL,
collided into the major polarities
(indicated with arrows),
and eventually filled the gapped PIL.
Figure \ref{fig:trigger}b
is the Ca image
of one particular collision event
in Panel (a)
(shown by a red arrow
in 08:00 UT),
where a positive patch
was advected and conflicted
with the preexisting negative polarity ({\sf N1}).
One may see that
a Ca brightening became more evident
above the collision site,
bridging over the both polarities
(indicated by arrows).
This Ca brightening suggests
magnetic reconnection
between the advected small magnetic patch
and preexisting polarities.
We found similar brightenings
all the way through
the formation
of the intrusive magnetic structure.
Therefore,
we expect that
the accumulation
of emerged/advected small-scale bipoles
is essential
for the formation of
the intrusive structure
in this AR.

\begin{figure}
  \includegraphics[scale=0.9,clip]{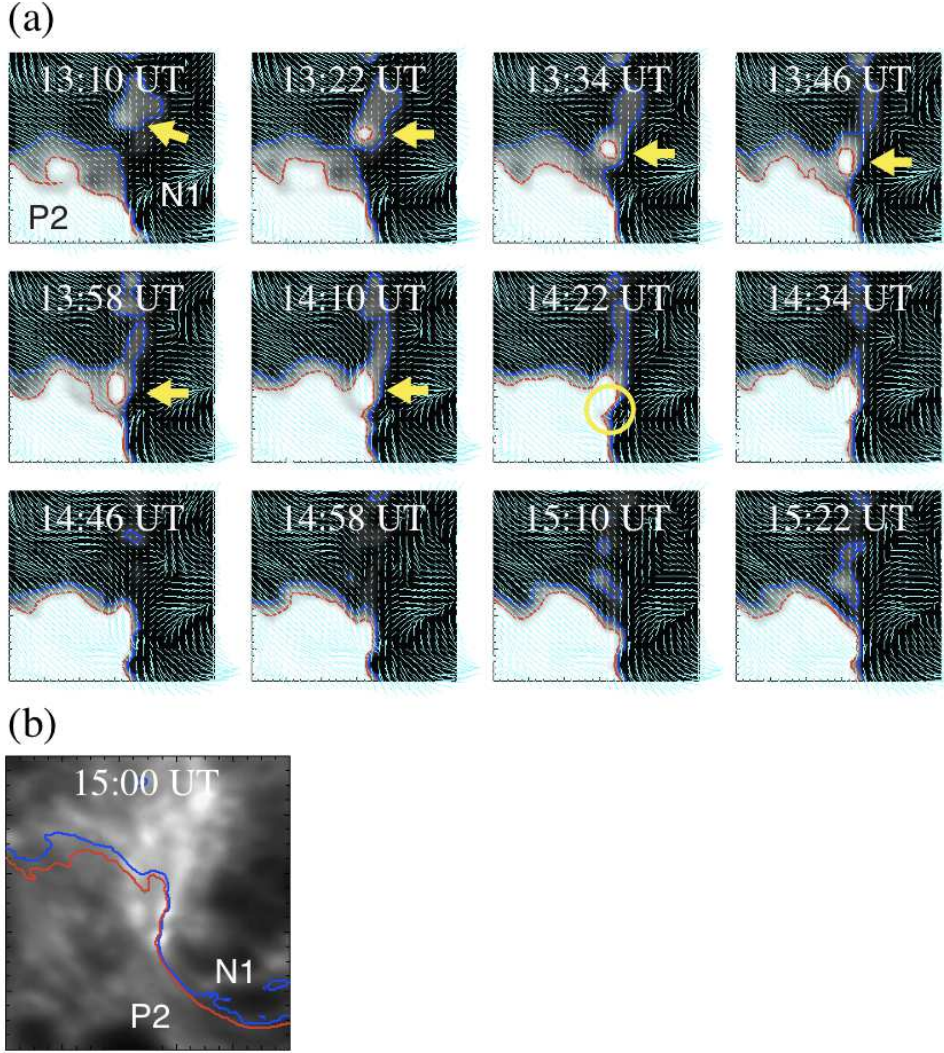}
  \caption{(a) HMI vector magnetogram
    during 13:10--15:22 UT.
    Grayscale shows the longitudinal field,
    while skyblue bars indicate the transverse field.
    Here the direction of the transverse fields
    are not indicated.
    Red and blue lines
    are the contour level of
    longitudinal component $=\pm 100\ {\rm G}$,
    respectively.
    The small positive patch
    indicated by arrows
    collides into the major positive patch {\sf P2}.
    At 14:22 UT,
    the transverse field in the circle
    connects
    the colliding positive patch
    and preexisting negative polarity
    {\sf N1}.
    (b) SOT Ca image
    around the PIL
    at 15:00 UT
    of the same day.
    Red and blue contours
    indicate the DN
    of the circular polarization
    $=\pm 20$, respectively.
   }
  \label{fig:bipole}
\end{figure}

Next, let us focus on
the collision event
that seems related to the
triggering of the M6.6 flare.
Before the flare occurrence,
there were many emergences
and collisions
of the patches.
One of the collision events
that possibly led to the M flare
is the case shown
in Figure \ref{fig:bipole}.
In Panel (a),
we plot HMI magnetograms
(longitudinal and transverse fields)
around the intrusive structure
for every 12 minutes
from 13:10 to 15:22 UT.
Here we do not indicate
the direction of the transverse fields
(transverse fields
are shown by bars
instead of arrows),
since the small-scale fields
we focus on
extend only for several pixels
in this map
and thus the orientation
of such fields
may have been smoothed out
in the inversion process.
At 13:10 UT,
a positive magnetic patch
started to emerge
and was then advected southward
(indicated with arrows).
At 14:22 UT,
the positive patch
merged into the main positive polarity
({\sf P2}: intrusive structure),
colliding against
the preexisting negative polarity ({\sf N1}).
The transverse fields
in the yellow circle
clearly connect
the advected positive patch
and the negative polarity,
which remains until 14:58 UT.
The transverse fields
connecting the both fields
indicate
the magnetic reconnection
between the positive flux of the advected patch
and the negative flux of the major polarity ({\sf N1}).
In this map,
however,
only six pixels (a size of a few Mm)
in the negative patch
(yellow circle)
were related to the reconnection.
Here,
the azimuth angle of the transverse fields,
averaged over the six pixels,
was measured to be $252^{\circ}$.
After the collision of the advected bipoles,
the SOT Ca image
at 15:00 UT
in Figure \ref{fig:bipole}b
reveals enhanced brightenings
around the PIL.

Here we summarize
the observations in this section.
The intrusive magnetic structure
(the flare-triggering region)
was formed through
the continuous accumulation
of the small-scale emerging bipoles.
When the positive patch collided against
the preexisting negative field,
we observed Ca brightenings
above the collision site,
which suggests a magnetic reconnection
between them.
One of the collision events
that possibly led to the triggering of the M6.6 flare
was observed between 14:00 to 15:00 UT,
February 13,
more than two hours before
the flare occurrence.
The azimuth angle
of the reconnected transverse field
was measured $252^{\circ}$,
while its spatial size was
only a few Mm.

\section{Comparison with Numerical Simulation:
  M6.6-class Flare
  \label{sec:flare}}

Observations
of the magnetic fields
related to the M6.6 flare
in AR 11158
indicate that
the geometrical relationship
between the overlying coronal arcade
and the intrusive magnetic structure
(the flare-triggering region),
namely,
the shear angle of the arcade
(80$^{\circ}$)
and the azimuth of the flare trigger
(270$^{\circ}$ to 300$^{\circ}$),
is consistent with
the RS model in \citet{kus12}.
Thus, in this section,
we provide a detailed comparison
of the observation
with the numerical results
of RS model.

The basic setup of the simulation
is described in \citet{kus12}.
Here we show the numerical result
with a shear angle
of the overlying linear force-free field (LFFF)
of 80$^{\circ}$
and an azimuth of the injected triggering flux
of 270$^{\circ}$,
which corresponds to the RS-type configuration.
Figure \ref{fig:simulation} shows
the temporal evolution of the simulation.
The left column shows
the selected magnetic field lines
and the current density in the central plane,
while the middle column
shows the magnetogram
with the shift of the field connectivity
(red: see Appendix \ref{app:footpoint})
and the current density
in the lower atmosphere (green).

In the initial state at $t=0.0$,
as of Panels (a) and (b),
one can see a highly-sheared coronal arcade
above the PIL ($y=0$).
When the bipolar flux was
injected from the bottom boundary,
Panels (d) and (e),
the current sheet {\sf P} was formed
between the injected flux
and the coronal fields,
and the primary reconnection
proceeded in the current sheet.
Due to the reduction
of the magnetic flux from the reconnection site {\sf P},
the force balance among the arcade fields
was partially lost,
which pulled the ambient arcade fields
into the center of the domain
and formed a twisted flux rope
via the secondary reconnection.
In Panel (j),
one may see that
the twisted flux rope {\sf T}
is detached and erupts upward,
creating
a vertical current sheet {\sf S} beneath.
(For the detailed illustration
of the field lines and current sheets,
the reader is referred to
Figure 4 of \citealt{kus12}.)
The footpoints of the fields lines
that connect to the flux rope {\sf T}
are labeled as {\sf F}
in Panels (h), (j), and (k),
while those connecting to
the postflare arcade
are shown as {\sf R}.

Compared with the numerical results,
SOT Ca images reveal
clear consistencies
throughout the whole evolution process.
At 15:25 UT
in Figure \ref{fig:simulation}f,
the preflare brightening {\sf P'}
above the PIL {\sf N1}--{\sf P2},
which was also shown
as the brightenings
in Figure \ref{fig:bipole}b,
was comparable to
the current sheet {\sf P}
between the injected triggering field
and the preexisting coronal arcade.
In Panel (i),
the flare two-ribbon {\sf F'}
in the eruptive phase
from 17:30 UT,
extending along the PIL
from the intrusive structure
(flare-triggering region),
was also well in accordance
with the footpoints
of the reconnected coronal fields {\sf F}
and of the postflare arcades {\sf R}
in Panel (h).
The flare ribbons
in the late phase in Panel (k),
however,
deviated from
the observation
of the Ca ribbons
in Panel (l).
In the simulation
both ribbons at $y=\pm0.2$
elongated along the PIL
to both positive and negative $x$-directions,
while in the actual Sun
the ribbons stopped their elongations
along the PIL from 17:35 UT.

It should be noted here that
the polarities in Panels (i) and (l)
show reversals,
possibly due to the strong Dopplershifts
that correspond to the M flare
\citep{fis12};
the black patches in the southern positive polarity
were actually positive rather than negative
and the white patches in the north
were negative.
The reversals were also observed
above the most distant sunspots
{\sf P1} and {\sf N2}.

There is also a difference
between the simulation and observation.
Here, in the numerical simulation,
the flare trigger was given as
a simple ``emerging flux''
with an RS configuration,
injected into the domain
from the bottom boundary.
On the other hand,
the observed flare trigger
(intrusive structure)
was created through
a continuous accumulation of
emerged/advected bipoles,
and the RS-component field
was supplied by
their magnetic reconnection
(see Section \ref{sec:trigger}).
These discrepancies
between the modeled and actual Sun
will be discussed
in the next section.

\section{Discussion
  \label{sec:discussion}}

In the previous sections,
we have investigated
the evolution of the magnetic fields
that caused the M6.6-class flare
in NOAA AR 11158,
from flux emergence
to the eventual eruption,
by spacecraft observations
and a comparison
with the numerical simulation.

AR 11158 was characterized
by its two major emerging fluxes
{\sf P1}--{\sf N1} and {\sf P2}--{\sf N2}
(Figure \ref{fig:arcade}).
In this quadrupolar region,
the central PIL
between {\sf N1} and {\sf P2}
was continuously sheared
by the counterclockwise motions
of the major polarities {\sf N1} and {\sf P2},
and, through this process,
the coronal arcade over this PIL
was highly sheared
(Figure \ref{fig:preflare}).
The observations indicate that
the M6.6 flare was triggered
by the intrusive magnetic structure
that appeared on this PIL.
The shear angle of the coronal arcade
and the azimuth of the intrusive structure
were measured to be
80$^{\circ}$ and 270$^{\circ}$--300$^{\circ}$,
respectively,
which corresponds to the RS-type model
in \citet{kus12}.
It was also found that
the intrusive structure
(the flare-triggering region)
was formed through
the continuous accumulation
of the emerged/advected bipoles
and their reconnections
with the preexisting negative field
(Figures \ref{fig:trigger} and \ref{fig:bipole}).
Based on the results above,
here we discuss some issues
on the magnetic structures
that correspond
to the flare onset
and the resultant eruption.

\subsection{Similarities and Differences
  between Modeled and Actual Sun}

In Section \ref{sec:flare},
we compared the observation
with the numerical results
of the RS type in \citet{kus12}.
We found that
the Ca preflare brightenings
above the triggering region
were highly consistent
with the numerical results
(Figures \ref{fig:simulation}e and f).
The brightenings
were explained as a current sheet
between the localized flare trigger
and the overlying coronal arcade.

The structure of
the flare ribbons
in the simulation
was also consistent
with the observation
(Figures \ref{fig:simulation}h and i).
However,
the simulated ribbon structure
deviated from that
in the actual Sun,
especially in the late phase
(Panels k and l).
In the computational domain,
the ribbons elongated
from the triggering field
($x=0$)
to both $x$-directions,
because the reconnections
among the coronal arcades
(given as linear force-free fields: LFFF)
occurred successively
along the PIL
(i.e., the $x$-axis).
On the other hand,
in the observation,
the ribbons
stopped their elongations
along the PIL,
possibly because
the region of the highly-sheared arcade
with 80$^{\circ}$
was localized around the flare trigger
(the intrusive structure)
and the shear outside this region
was not so high as $\sim 80^{\circ}$,
which inhibited the successive reconnections
of the coronal arcades
propagating along the PIL.
In other words,
the non-uniformity of the shear angle
in the AR,
or, the effect of
non-linear force-free fields (NLFFF),
limited the elongation
of the flare two-ribbons.
This story is well in line with
the simulation results
of the RS type with different shear angles;
the flare eruption was observed
in the highly-sheared case,
while it failed
in the weakly-sheared case
(compare the cases with azimuth of 270$^{\circ}$
in Figure 2 of \citealt{kus12}).

In addition,
there was an important difference
between the simulation and the observation.
In the simulation,
the flare trigger was given
as a simple ``emerging flux''
from the bottom boundary.
However,
the flare trigger
of the M6.6 event in AR 11158
was formed through
a series of bipolar emergences
and their reconnections
with the preexisting field.
This discrepancy may suggest
the possibility that
the flare-triggering region
in the actual Sun
can be achieved through
a variety of dynamic processes,
while the essential magnetic configuration
can be classified into
the two simple categories,
OP and RS.

Here we comment on the contrast
of the time scale
between the simulation 
(Figure \ref{fig:simulation} left and middle columns)
and the observation
(right column).
If we take $0.4\ {\rm s}$
for a normalizing unit of the time
following
\citet{kus12},
the entire calculation spans
only $16\ {\rm s}$,
which is three orders of magnitude
shorter than the observed time range
of the M6.6 flare.
In the simulation,
the emergence speed
of the injected triggering flux
is made faster than
the actual flux emergence
in the Sun
($\sim 1\ {\rm km\ s}^{-1}$)
by two orders,
in order to reduce the computation time
and focus mainly on behaviors
after the flux injection
\citep[see also Section 4.2 of][]{kus12}.
Also,
the electrical resistivity in the simulation
is set much larger,
which may drastically speed up
the magnetic reconnection
between the injected flux and the overlying arcade.
However,
since the emergence in the simulation
is much slower than the Alfv\'{e}n velocity,
the results after the injection
do not depend much on the emergence speed.
Plus,
although the reconnection proceeds much faster
in the computation,
we believe that
the topological evolution of the magnetic structure
in the nonlinear phase
is properly solved
because the reconnection time is longer than
the Alfv\'{e}n time (dynamical time scale)
and is much shorter than
the diffusion time.

\subsection{Large-scale Magnetic System,
  Subsurface History,
  and Successive Flares
  \label{sec:large-scale}}

In its birth of AR 11158,
we observed two major emerging fluxes
{\sf P1}--{\sf N1} and {\sf P2}--{\sf N2}
(see Figure \ref{fig:arcade}).
The motions of the quadrupole
were described as
two inner polarities {\sf N1} and {\sf P2}
moving rapidly
as if they merged into
rather-stable outer polarities
{\sf N2} and {\sf P1},
respectively.
The relative motion of {\sf N1} and {\sf P2}
built a highly-sheared PIL
in the center of the AR
and thus the magnetic free energy
was stored in the corona.
Finally, the M6.6 flare,
which had an RS configuration,
broke out above this PIL.

From the point of view
that the flare is a relaxation process
leading to a lower energy state,
along with
the proper motions of the quadrupole,
this AR is strongly reminiscent
of the history that
a single flux tube
emerged from the deeper convection zone
split into two
by some perturbations
and these split tubes
appeared at the neighboring locations
at the visible surface
\citep[see also][]{chi13}.
Figure \ref{fig:tube}a illustrates
this concept.
Here the two flux tubes
that appear at the photosphere
({\sf P1}--{\sf N1} and {\sf P2}--{\sf N2})
share the common roots
below the surface.
The relative motion of {\sf N1} and {\sf P2}
and the resultant M flare
can be explained
by the rising
of this large-scale magnetic system
as a whole.
Therefore,
to understand the cause of flares,
it is surely important to investigate
the subsurface magnetic fields by
helioseismology \citep[e.g.,][]{ilo11,tor13}
and numerical approaches \citep[e.g.,][]{tor10,tor12}.

The flare simulation
in Section \ref{sec:flare}
showed that
the filament eruption
of the M flare on February 13, 2011,
was caused through
the collapse of the coronal arcades
(Figure \ref{fig:simulation}j).
This situation is in accordance with
the tether-cutting model
suggested by \citet{moo01}
and the magnetic structure
is consistent with
the classical CSHKP model
\citep{car64,stu66,hir74,kop76}.
Also,
the numerical results
predict that,
as the filament erupted into the space,
there remained a postflare arcade
beneath the detached filament
({\sf R} in Figure \ref{fig:simulation}j),
which is a remnant
of the reconnected coronal arcades.
\citet{liu12} found that
the photospheric horizontal fields
increased drastically
after the M flare.
They discussed in favor
of the tether-cutting reconnection
that the upper part of the reconnected arcades
erupted outward,
while the lower part contributed 
to enhance the photospheric fields.
The lower fields may be the same
as the postflare arcade
in the simulation results
of our study.

After the M6.6 event,
the newly created postflare arcade
connecting {\sf N1} and {\sf P2}
was sheared further
by the continuous photospheric motions.
We believe that the continuous motions
were also caused
by the large-scale flux system,
namely, the emergence of the spilt tubes
(Figure \ref{fig:tube}a).
The highly-sheared coronal arcade
was thus formed again
along the same PIL
between {\sf N1}--{\sf P2},
which may be the overlying arcade
in the X2.2 flare
that occurred on February 15.
\citet{wan12} also observed
the enhancement
of the photospheric horizontal fields
after the X flare,
explaining that
the X event was similarly caused
by the tether-cutting reconnection
\citep[see also][]{sun12}.

A series of flares
along the central PIL
in AR 11158,
including X2.2, M6.6,
and many C-class events,
indicate that
(1) successive flares
in a single AR
are the way to relax
the continuous shearing
of the magnetic fields
by the photospheric motions,
or, by the large-scale flux emergence
of the whole magnetic system,
(2) the timing of each flare
is determined
by the appearance
of a flare trigger,
and (3) the magnitude
of each event
depends on
the conditions of the event,
e.g., the amount of coronal flux
aligned on the PIL,
the shear angle
of the coronal arcades,
and the azimuth angle
of the triggering field.
And,
the large-scale magnetic system
emerged from the subsurface layer
is thus important
for the successive flares.

\subsection{Multiscalability of the Magnetic Systems}

As was seen in the observations,
magnetic structures in AR 11158
extended widely over
multiple spatial scales,
ranging from the entire length
of the AR of the order of 100 Mm
to the size of the RS-component flux
of a few Mm.
Also the time scale distributed
from days to hours.

Figure \ref{fig:tube}a shows
the largest-scale structure
in this AR,
i.e.,
the double bipolar systems
{\sf P1}--{\sf N1} and {\sf P2}--{\sf N2}.
The size of each emergence
extended more than 40 Mm
and the development
lasted for several days.
As was discussed
in Section \ref{sec:large-scale},
the proper motions of the quadrupole
can be explained
as the emergence of the two split tubes
that share the common roots.
If this is the case,
the series of flares
above the PIL
can be understood
as the process
that the split flux tubes
recovered their original shape
(a single tube)
when they rose into the corona,
where the gas pressure
is less dominant.

The M6.6-class flare
analyzed in this study
was found to be triggered
by the smaller-scale magnetic field.
Figure \ref{fig:tube}b shows
a closeup of the PIL
between {\sf N1} and {\sf P2},
where these polarities sheared
the coronal arcade
by their relative motions.
It was found that,
the intrusive magnetic structure
of $\sim 5\ {\rm Mm}$,
localized in the center of the PIL,
was a possible flare trigger.
From Figures \ref{fig:preflare}a--c,
one can see the lifetime
of the intrusive structure
is less than a half day,
or about 10 hours.

The formation of the intrusive structure
on the PIL
(the flare-triggering region)
was caused by
much smaller-scale events.
Figure \ref{fig:tube}c illustrates
the formation process
of the intrusive structure.
First,
small-scale bipoles
of the size of a few Mm
emerged continuously
in the middle of the gapped PIL.
Some of the positive patches
were then advected to the west
and finally collided
into the preexisting negative field {\sf N1}
(see Figure \ref{fig:trigger}a).
The time scale of each advection/collision
was about, or less than 1 hour.
At the same time,
we observed brightenings
in the Ca image
(Figure \ref{fig:trigger}b),
which indicates the reconnection
between the advected patch
and the preexisting fields.
Magnetic flux of RS component
was thus achieved.

The multiscalability
of the magnetic fields
both in spatial and temporal scales
indicates that,
even in the large-scale events
like flares and resultant eruptions,
smaller-scale physics
plays an important role
in driving the entire system.
This may also indicate that, 
for the flare prediction,
we have to deal with
the short-lived flare trigger
of the time scale of a few hours.
Therefore,
to completely understand
the nature of flares,
further studies that simultaneously
handle
the small-scale/short-term flare trigger
and the large-scale/long-term evolution
of the entire AR
are required.

\subsection{Categorization of Flare-Productive ARs}

As discussed above,
the large flares in AR 11158
occurred mainly on the continuously-sheared PIL,
which is formed by
a relative motion
of two major emerging bipoles.
Our explanation of the flares
is that they are the relaxation process
of a single flux tube,
which has been split into two tubes
by some dynamics in the convection zone,
restoring its original geometry
(Figure \ref{fig:tube}a).
And one of these flares,
the M6.6 event,
was found to be consistent
with the RS-type model.

On the contrary,
NOAA AR 10930,
also a flare-productive region
sourcing several X-class events,
basically consisted of
a single bipole
\citep{kub07}.
The long-term evolution
in this AR was that
the southern positive polarity
passed by the large, northern negative polarity
from west to east,
while the positive polarity itself
rotated counterclockwise,
forming a highly sheared PIL.
\citet{kus12} focused on this PIL
and found that
the X3.4-class flare
on 2006 December 13
can be explained by the OP model.
The possible scenario
of this AR is that,
through a series of flares,
the twisted flux tube
forming this region
expelled its internal helicity (or shear)
into the space
for relaxation
to a lower energy state.

From these examples,
we can see that
there exist some scenarios
for the flare-productive ARs.
The flux tubes of
complex, multipolar regions
such as AR 11158
may have suffered severe interruptions
during their ascents
in the convection zone.
And thus,
they may try to restore
their original shapes
via flares
as a mean of relaxation
\citep[see also][]{poi13}.
Meanwhile,
the tubes of
simpler, perhaps bipolar regions
such as AR 10930
may have been tightly twisted
under the visible surface.
As they appear
into the atmosphere,
the twisted tubes
may lighten up
their helicities
in the form of flares.

The above categorizations of the ARs,
along with the triggering mechanisms
of OP and RS,
can be used
to classify the flaring events.
However,
it is not clear
whether complex, multipolar regions
are in favor of the RS model,
and simple, bipolar regions
are for OP.
Statistical studies may help
to understand the relationship
among these elements.

\section{Summary
  \label{sec:summary}}

NOAA AR 11158
was a flare-productive region,
mainly composed of two large-scale emerging fluxes.
Because of the relative motions
of the two inner polarities,
a highly sheared PIL
was created
in the center of this AR,
which then sheared
the coronal arcade
lying over the PIL.
Based on the observations,
we here interpret that
the M6.6-class flare
on 2011 February 13
was triggered
by a localized magnetic region
that had an intrusive topology,
the positive polarity
penetrating into the negative side.
The geometrical relationship
between the flare trigger
and the coronal arcade
was in favor of the RS scenario
of \citet{kus12}.
The shear angle
of the arcade
and the azimuth angles
of the flare trigger
were measured to be
$80^{\circ}$ and $270^{\circ}$--$300^{\circ}$,
respectively,
which agrees with the RS model.
We found that
the development of the flare-triggering region
was due to a series of small-scale emerging bipoles,
a few hours before the flare occurrence.
Through magnetic reconnections between
advected positive patch
and the preexisting negative field,
magnetic flux of the RS component
was thus achieved.
Finally,
after the triggering region was formed,
the M flare
started at around 17:30 UT.
These results indicate that
a wide spectrum of magnetic systems,
ranging from the large-scale evolution of the entire AR
to the small-scale accumulation of magnetic elements,
is altogether involved
in a flare activity.




\acknowledgments

The authors would like to
thank the anonymous referee
for improving the paper.
This work was supported
by a Grants-in-Aid for Scientific Research (B)
``Understanding and Prediction of Triggering Solar Flares''
(23340045, Head Investigator: K. Kusano)
from the Ministry of Education, Science, Sports,
Technology, and Culture of Japan.
{\it Hinode} is a Japanese mission
developed and launched by ISAS/JAXA,
with NAOJ as domestic partner
and NASA and STFC (UK) as international partners.
It is operated by these agencies
in co-operation with ESA and NSC (Norway).
The authors would like to thank
the {\it SDO} team for distributing HMI and AIA data.
The numerical simulation was conducted
on the Earth Simulator
in Japan Agency for Marine-Earth
Science and Technology (JAMSTEC).






\appendix

\section{Initial and Boundary Conditions
  for the NLFFF Extrapolation
  \label{app:nlfff}}

To obtain the initial and boundary conditions
for the NLFFF extrapolation,
we decompose the magnetic field
normal to the solar surface $B_{n}$
to the averaged component $\bar{B}_{n}$
and the undulate component 
$b=B_{n}-\bar{B}_{n}$.
Assuming a periodic condition
for the horizontal coordinates,
we calculate a potential magnetic field
for $b$ in terms of
the Fourier transform
\begin{eqnarray}
  \mbox{\boldmath $B$} = \sum_{m,n}
  \tilde{\mbox{\boldmath $b$}}_{(m,n)} \exp{(ik_{m} x + ik_{n} y - |k|z)},
\end{eqnarray}
where $k_{m} = 2\pi m/L_{x}$,
$k_{n} = 2\pi n/L_{y}$,
$|k| = \sqrt{k_{m}^2+k_{n}^2}$,
$m = -n_{x}/2, \cdots, -1, 1, \cdots, n_{x}/2$,
and 
$n = -n_{y}/2, \cdots, -1, 1, \cdots, n_{y}/2$,
respectively.
The complex Fourier coefficient vector 
$\tilde{\mbox{\boldmath $b$}}_{(m,n)}$
consists of
$\tilde{b}_{x(m,n)} = -i k_{m} \tilde{b}/|k|$,
$\tilde{b}_{y(m,n)} = -i k_{n} \tilde{b}/|k|$, 
and $\tilde{b}_{z(m,n)}=\tilde{b}$ that is
a Fourier transform of $b$.

In order to compensate
the excess magnetic flux $\bar{B}_{n} S_{b}$
on the bottom boundary,
we add a small bias field
$-\bar{B}_{n} S_{b}/S_{a}$
to the normal component of 
magnetic field on all the lateral and top boundaries,
where $S_{b}$ and $S_{a}$ are
the area of the bottom boundary
and the total area
of the lateral and top boundaries,
respectively.
Then, we derive the potential field
$\mbox{\boldmath $B$}=-\nabla \Phi$,
solving the Poisson equation $\nabla^2 \Phi = 0$ numerically.
This potential field is used
as the initial condition
for the NLFFF calculation.
Incrementally imposing the transverse components
of the observed vector magnetogram
onto the bottom boundary, 
the NLFFF is iteratively calculated based
on the MHD equation.
The normal components
on all the boundaries are fixed,
while the transverse components
on the lateral and top boundaries 
are solved
under the condition
that these boundaries are 
perfectly conductive.

\section{Shift of the Field Connectivity
  \label{app:footpoint}}

\begin{figure}
  \includegraphics[scale=1.1,clip]{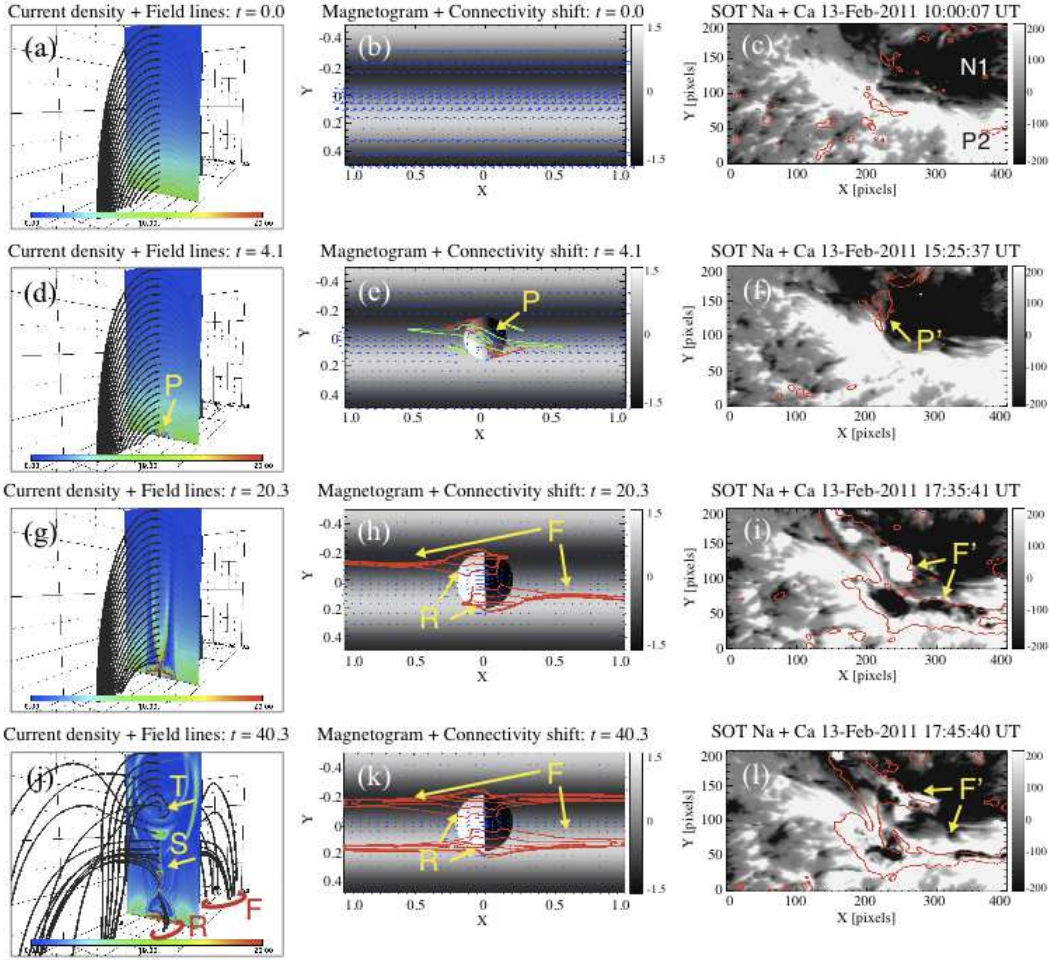}
  \caption{(Left) Numerical results
    of the RS-type simulation.
    Selected field lines
    passing through the vertical axis $x=y=0$
    and current density
    in the vertical plane $x=0$
    are shown.
    The preflare brightening {\sf P},
    twisted flux rope {\sf T},
    current sheet {\sf S},
    flare two-ribbons {\sf F},
    and postflare arcade {\sf R}
    are also indicated.
    (Middle) Simulation results
    showing photospheric vertical
    and horizontal fields
    (grayscale and blue arrows).
    Red contours present
    the shift of the field connectivity
    (see Appendix \ref{app:footpoint}).
    Green contours in Panel (e)
    are the current density
    in the lower atmosphere,
    averaged over $0.05\le z \le 0.15$.
    (Right) SOT Na Stokes-V/I images
    overplotted by the Ca intensity
    (red contour).
    The preflare brightening {\sf P'}
    and flare two-ribbons {\sf F'}
    are also indicated.
  }
  \label{fig:simulation}
\end{figure}

\begin{figure}
  \includegraphics[scale=0.7,clip]{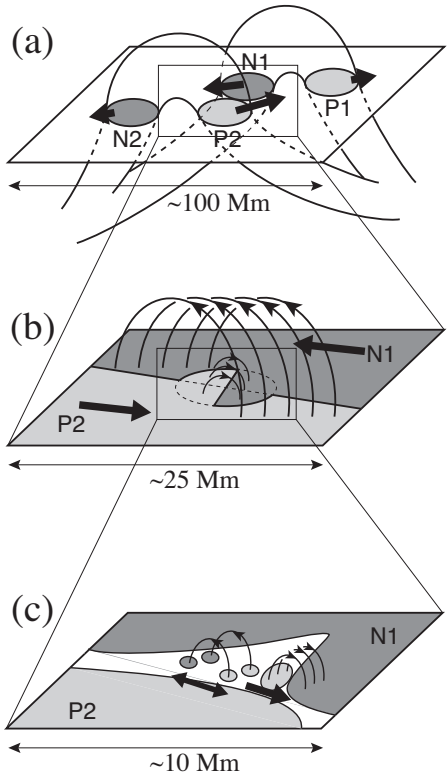}
  \caption{
    Schematic illustration
    of the magnetic structures
    in multiple scales
    that are involved
    in the M6.6-class flare
    in NOAA AR 11158.
    (a) Ellipses
    on the plane (photosphere)
    indicate the two major bipoles,
    {\sf P1}--{\sf N1}
    and {\sf P2}--{\sf N2}.
    Lighter and darker shadows
    mean the positive and negative polarities,
    respectively.
    Tubes above and below the photosphere
    show the expected flux tubes
    that compose this AR.
    (b) Closeup of the PIL
    between {\sf N1} and {\sf P2}.
    Relative motions in both sides
    of the PIL shear the overlying coronal arcade,
    while, in the core of the PIL,
    flare-triggering region
    (intrusive structure)
    with the RS-component flux appears.
    (c) The formation of the triggering region
    is illustrated.
    Small-scale bipoles emerge
    in the gapped PIL,
    and the positive patch is advected
    to collide into {\sf N1} polarity,
    forming RS flux
    through magnetic reconnection.
   }
  \label{fig:tube}
\end{figure}

In the middle column
of Figure \ref{fig:simulation},
we plot with red contours
the shift of the field connectivity
from the initial linear force-free condition.
From each point of the bottom boundary,
$\mbox{\boldmath $x$}_{0}$,
the end point of the field line at the time $t$
is traced,
which is denoted as
$\mbox{\boldmath $x$}_{1}(\mbox{\boldmath $x$}_{0};t)$.
In this map,
the red contour indicates
the shift of the end point at $t$
from the initial force-free state,
$|\mbox{\boldmath $x$}_{1}(\mbox{\boldmath $x$}_{0};t)
-\mbox{\boldmath $x$}_{1}(\mbox{\boldmath $x$}_{0};0)|$,
at each location $\mbox{\boldmath $x$}_{0}$.
If this value is larger,
the field line traced
from that position
has reconnected
with another field line
which was initially located farther.




\clearpage

\clearpage

\clearpage

\clearpage

\clearpage

\clearpage

\clearpage

\clearpage

\end{document}